\documentclass[aps,twocolumn,nofootinbib,superscriptaddress]{revtex4-1}
\usepackage{amssymb}
\usepackage{amsmath}
\usepackage{amstext}
\usepackage{amsfonts}
\usepackage{bbold}
\usepackage{bm}
\usepackage{graphics}
\usepackage{mathtools}
\usepackage{lmodern}
\usepackage{graphicx}
\usepackage[capitalise]{cleveref}
\usepackage{xcolor}

\def\gsim{ \lower .75ex \hbox{$\sim$} \llap{\raise .27ex \hbox{$>$}} }
\def\lsim{ \lower .75ex \hbox{$\sim$} \llap{\raise .27ex \hbox{$<$}} }

\begin{document}

\title{A non-perturbative test of consistency relations and their violation} 

\author{Angelo~Esposito}
\affiliation{Theoretical Particle Physics Laboratory (LPTP), Institute of Physics, EPFL, 1015 Lausanne, Switzerland}
\author{Lam~Hui}
\affiliation{Department of Physics, Center for Theoretical Physics, Columbia University, 538W 120th Street, New York, NY, 10027, USA}
\author{Roman~Scoccimarro}
\affiliation{Center for Cosmology and Particle Physics, Department of Physics, New York University, NY 10003, New York, USA}

\begin{abstract}
In this paper, we verify the large scale structure consistency relations 
using $N$-body simulations,
including modes in the highly non-linear regime. 
These relations (pointed out by Kehagias \& Riotto and Peloso \& Pietroni) follow from the symmetry of the dynamics under a shift of the Newtonian potential by a constant and a linear gradient, and predict the absence of certain poles in the ratio between the (equal time) squeezed bispectrum and power spectrum. The consistency relations, as symmetry statements, are exact, but have not been previously checked beyond the perturbative regime. Our test using $N$-body simulations not only offers a non-perturbative check, but also serves as a warm-up exercise for applications to observational data. A number of subtleties arise when taking the squeezed limit of the bispectrum---we show how to circumvent or address them. 
An interesting by-product of our investigation is an explicit demonstration that the linear-gradient-symmetry is unaffected by the periodic boundary condition of the simulations. 
Lastly, we verify using simulations that the consistency relations are
violated when the initial 
conditions are non-gaussian (of the local $f_{\rm NL}$ type). 
The methodology developed here paves the way for constraining primordial non-gaussianity
using large scale structure data, including (numerous) highly non-linear modes that are 
otherwise hard to interpret and utilize. 
\end{abstract}

\maketitle


\section{Introduction}
\label{intro}

One of the key questions in modern cosmology concerns the initial condition of the universe. Are the primordial fluctuations consistent with what one would expect from single-field inflation? Or do they arise from a scenario in which
additional light fields, besides the inflaton, play an important role?
Or more radically, is some mechanism other than inflation at work? 

The standard approach to answering these questions is to work with
data in the linear or quasi-linear regime where perturbation theory
can be relied upon to give reliable predictions.
Modes in the non-linear regime (for instance, with momentum $k \, \gtrsim \, 0.2$ $h$/Mpc in large scale structure data) are not utilized, even though they are 
abundant and measured with high precision. 

The consistency relations offer an interesting alternative, where some of the information hidden in the nonlinear regime can be brought to light. First pointed out by Maldacena~\cite{Maldacena:2002vr}, consistency relations
connect a squeezed $(N+1)$-point correlation function (squeezing means one of the momentum legs is soft) to an $N$-point function (see also \cite{Creminelli:2004yq,Cheung:2007sv}).
More recent work pointed out additional consistency relations coming from
new symmetries,
clarified the assumptions behind consistency relations and emphasized
their exact, non-perturbative nature, analogous to soft theorems in
high energy physics 
\cite{Creminelli:2012ed,Hinterbichler:2012nm,Assassi:2012zq,Kehagias:2012pd,Hinterbichler:2013dpa,Goldberger:2013rsa,Hui:2018cag}.
The non-perturbative nature of consistency relations is a mere curiosity 
for the microwave background since its fluctuations are small and linear, but
becomes very interesting for large scale structure.
Kehagias \& Riotto and Peloso \& Pietroni~\cite{Kehagias:2013yd,Peloso:2013zw} pointed out the
relevant large scale structure consistency relations. It can be shown
that of the infinite tower of general relativistic consistency
relations \cite{Hinterbichler:2013dpa}, two
has non-trivial Newtonian, sub-Hubble limits
\cite{Creminelli:2013mca,Horn:2014rta}.

The study of large scale structure concerns, at a minimum, the following quantities: the mass fluctuation $\delta$, the peculiar velocity $\vec v$  and the gravitational potential $\Phi$. (One can further expand this list to include the galaxy count fluctuation $\delta^g$ and the galaxy peculiar velocity  $\vec v^{\,g}$.
The symmetries discussed below apply to them as well, where $\delta^g$ and $\vec v\, {}^g$ transform in the same way as $\delta$ and $\vec v$ do, see e.g.~\cite{Kehagias:2013rpa,Kehagias:2015tda})
The dynamics of (sub-Hubble) fluctuations exhibits two non-linearly
realized symmetries in a matter + cosmological constant
universe.\footnote{The split into two separate symmetries here follows the
  discussion of \cite{Horn:2015dra}.}
One is a constant shift in the gravitational potential:
\begin{align}
\label{shiftSymm}
\Phi \rightarrow \Phi + c
\end{align}
where $c$ is independent of space but possibly a function of time. The other involves adding a linear gradient to the gravitational potential, together with a transformation of the spatial coordinates and the velocity field~\cite{Horn:2014rta}:
\begin{align}
\label{lineargradientSymm}
\vec x \rightarrow \vec x + \vec n\, , \;\;
\Phi \rightarrow \Phi - ({\cal H} \vec n^\prime + \vec n^{\prime\prime}) \cdot \vec x 
\, , \;\;
\vec v \rightarrow \vec v + \vec n^\prime
\end{align}
where $\vec n$ is independent of space but a function of time.
Here $^\prime\equiv\partial/\partial\eta$ is the derivative with
respect to the conformal time $\eta$, and $\mathcal{H}\equiv
a^\prime/a$ is the comoving Hubble parameter, with $a$ being the scale
factor. The above is a symmetry of the large scale structure dynamics
for $\vec n$ having any time dependence, but the adiabatic mode
condition \cite{Weinberg:2003sw} dictates that $\vec n$ must match the time-dependence of the
linear growth factor, and
likewise $c$ should match the corresponding time-dependence of the
gravitational potential 
(see discussions in \cite{Horn:2015dra,Hui:2018cag} and point 2 below).

The consistency relations corresponding to a shift of the gravitational potential by a constant
and by a linear gradient are respectively:
\begin{align}
\label{CR1}
\lim_{\vec q\to0}q^2\frac{\langle \delta_{\vec q}\,\delta_{\vec k_1}\cdots \,\delta_{\vec k_N}\rangle^{c^\prime}}{P_\delta(q)}=0\,,
\end{align}
and
\begin{align}
\label{CR2}
\begin{split}
&\lim_{\vec q\to0}\vec\nabla_{\!q}\left[q^2\frac{\langle \delta_{\vec q}\,\delta_{\vec k_1}\cdots \,\delta_{\vec k_N} \rangle^{c^\prime}}{P_\delta(q)}\right]  \\
&=-\sum_{a=1}^N\frac{D(\eta_a)}{D(\eta)}\vec k_a\langle\delta_{\vec k_1}\cdots \,\delta_{\vec k_N}\rangle^{c^\prime}\, ,
\end{split}
\end{align}
where $P_\delta(q)$ is the mass power spectrum, $D$ is the linear
growth factor and $c^\prime$ denotes the connected correlator with the 
overall $\delta$-function removed. The time dependence is as follows:
the soft mode $\vec q$ is at time $\eta$ (likewise for $P_\delta(q)$) while
the hard mode $\vec k_a$ is at time $\eta_a$. 
Several comments are called for on these two consistency relations. 

\vspace{1em}

\noindent 1. The consistency relations are in general of an unequal time form.
In this paper, we focus on the equal time limit, in which case
the right hand side of the Eq. \eqref{CR2} vanishes. 
Thus, the content of the consistency relations is simple, that
the {\it equal time} correlator
\begin{align}
\label{nopoles}
{\langle \delta_{\vec q}\, \delta_{\vec k_1} \dots \delta_{\vec k_N} \rangle^{c'} \over P_{\delta} (q)}
\quad {\rm has \,\, no\,\,} q^{-2} \,\, {\,\rm pole} \,\, 
\& {\rm \,\, no\,\,}
q^{-1} \,\, {\rm pole}
\end{align}
in the $\vec q \rightarrow 0$ limit. The lack of a $1/q^2$ pole
follows from the shift symmetry, and the lack of a $1/q$ pole follows
from the linear gradient symmetry. That this statement is correct
(for gaussian initial conditions) is easy to check
in perturbation theory (see e.g.~\cite{Peloso:2013zw,Kehagias:2013yd,Horn:2014rta}). But the consistency relations, as
symmetry statements, are expected to be stronger than this.
What we wish to accomplish in this paper is to test this statement in
the non-perturbative regime using $N$-body simulations (i.e. with the
hard momenta ${\vec k_a}$'s on nonlinear scales).\footnote{We focus on the equal time correlator largely for simplicity.
There is also a practical reason for doing so: that
the $1/q$ pole associated with the unequal time contributions (i.e. the
right hand side of Eq.~\eqref{CR2}) is naturally suppressed in
observational data. Recall
that the unequal times refer to the times of the hard
modes ($\eta_1$ for $\vec k_1$, $\eta_2$ for $\vec k_2$ and so on);
the hard modes are 
by definition short wavelength perturbations
which also means their separation in time cannot be too big---keep
mind that observational data are confined to the light cone.
One can see from Eq.~\eqref{CR2} that if the $\eta_a$'s are close to
each other, one is almost summing the $\vec k_a$'s which yields zero.
Nonetheless, it is worth asking how big of a $1/q$ pole one
might inadvertently generate by measuring a correlator averaged
over some survey volume, which inevitably spans a range of redshifts.
Some care in defining the average might be useful to ensure
it is negligible. It is also worth noting that the unequal-time
contributions do not generate a $1/q^2$ pole.
A $1/q^2$ pole can only appear with certain primordial
non-gaussianities
(see point 6 below).
}

\vspace{1em} 

\noindent 2. It should be emphasized that the consistency relations are not statements
merely about a strictly vanishing $\vec q$. Indeed, an exact $\vec q=0$ mode is not even
observable. Rather, the consistency relations are statements about the absence of
certain divergences as $\vec q$ is taken to be smaller and smaller, such as 
\eqref{nopoles}. This is why the so called adiabatic mode condition is crucial
\cite{Weinberg:2003sw,Mirbabayi:2016xvc}. 
This condition ensures that the symmetry in question, which in general originates as 
a gauge redundancy, generates a $\vec q=0$ mode that is smoothly connected to
a physical mode of a small but finite $\vec q$. 
For more discussions on this point, see \cite{Hui:2018cag}.

\vspace{1em}

\noindent 3. The consistency relations Eqs.~\eqref{CR1} and
\eqref{CR2} take a particularly 
simple form in 
Lagrangian space where the corresponding ``right hand side'' vanishes even if
the hard modes are at unequal times. See \cite{Horn:2015dra} for a discussion.

\vspace{1em}

\noindent 4. The consistency relations take essentially the same form even in redshift space, as pointed out by \cite{Creminelli:2013poa}.
This means they can be profitably applied in galaxy surveys where the line-of-sight direction is
almost always in redshift space.

\vspace{1em}

\noindent 5. There is the question of how galaxy biasing affects the consistency relations.
As mentioned above, the relevant symmetries remain good symmetries even
for the dynamics of galaxies (which can form, merge and so on).\footnote{Galaxy dynamics is of course different from mass dynamics: 
mass conservation is replaced by galaxy number density evolution
that has a source (or sink) term; galaxies are subject to forces beyond
gravity. The key observation is that as long as these new terms/forces 
depend only on mass/galaxy density, velocity gradients (or velocity
difference between different species) and second derivatives of the
gravitational potential (tidal forces), the symmetries espoused in
Eqs.~\eqref{shiftSymm} and \eqref{lineargradientSymm} hold. 
For instance, it is crucial the new forces on a galaxy do not depend on 
the absolute velocity, i.e. some form of equivalence principle
(see point 6 below).
There's an additional requirement: that the squeezed momentum $\vec q$
must be sufficiently soft, that on that scale, gravity dominates (even
though for the hard momenta $\vec k$'s, the dynamics can be complicated).
See \cite{Peloso:2013spa,Horn:2014rta} for further discussions.
}
Thus, the consistency relations Eqs.~\eqref{CR1} and \eqref{CR2}
remain valid even if the hard modes $\delta_{\vec k_1}\, , \dots , \,\delta_{\vec k_N}$ are replaced
by galaxy density fluctuations $\delta^g {}_{\vec k_1} \, , \dots , \,\delta^g {}_{\vec k_N}$.
The soft mode $\delta_{\vec q}$ can be replaced by $\delta^g {}_{\vec q} / b^{g}$ where
$b^g$ is the galaxy bias; likewise $P_\delta (q)$ can be replaced by $P_{\delta^g} (q) / b^g {}^2$.
In the soft limit, $b^g$ is expected to be a constant\footnote{This
  holds if the initial conditions were gaussian, an assumption that
  goes into the derivation of the consistency relations themselves. Or
  more precisely, this assumes single-field or single-clock initial
  conditions. See discussion in point 6 below.},  
and thus the consistency relations Eqs.~\eqref{CR1} and \eqref{CR2} are modified in a simple way.
The equal time version \eqref{nopoles} in fact takes the same form i.e.
the equal time correlator:
\begin{align}
\label{nopolesG}
{\langle \delta^g {}_{\vec q} \delta^g {}_{\vec k_1} \dots \delta^g {}_{\vec k_N} \rangle^{c'} \over P_{\delta^g} (q)}
\; {\rm has \,\, no\,\,} q^{-2} \,\, {\,\rm pole} \,\, 
\& {\rm \,\, no\,\,}
q^{-1} \,\, {\rm pole}
\end{align}
in the $\vec q \rightarrow 0$ limit.

\vspace{1em}

\noindent 6. Two important assumptions go into deriving the consistency relations.
One is the equivalence principle, that on sufficiently large scales---i.e. $\vec q \rightarrow 0$---all objects fall at the same rate (whereas on small scales, different objects can be subject to
different forces, such as pressure forces, etc). See
\cite{Creminelli:2013nua,Horn:2014rta} for a discussion.
The other important assumption, which we focus on in this paper, is gaussian initial conditions.
More precisely, it is the assumption that in the squeezed limit, the primordial connected 
$N$-point function vanishes for $N>2$, something that follows from single-clock inflation.\footnote{
The primordial consistency relations can be expressed as the vanishing
of the squeezed $N$-point function if one accounts for the fact that
the metric fluctuations enter into the definition of physical
momenta. See \cite{Tanaka:2011aj,Pajer:2013ana,Bravo:2017gct,Bravo:2017wyw} for a discussion.
}
From the point of view of initiating cosmological $N$-body simulations, imposing
gaussian initial conditions is sufficient to guarantee the validity of the consistency relations 
stated above, and this is what we adopt in this paper.
It is not surprising that the consistency relations, or the precise form they take, are sensitive to 
initial conditions, since the symmetries underlying them are non-linearly realized or spontaneously broken---in other words, exactly how the initial conditions, or the ``vacuum'', breaks the symmetries in question dictates the form of the consistency relations \cite{Hui:2018cag}. 
Examples that violate the stated consistency relations generally involve extra light fields during inflation, for instance 
the curvaton, a spectator scalar that dominates the curvature fluctuations~\cite{Lyth:2006gd,Dvali:2003em,Kofman:2003nx}.\footnote{Ultra-slow-roll inflation, while strictly a single field model, has essentially an extra clock
due to the importance of what normally would be discarded as the
decaying mode. 
See \cite{Namjoo:2012aa,Martin:2012pe,Finelli:2017fml,Hui:2018cag}.}
The curvaton (or modulated-reheating) model motivates initial
conditions of the local $f_{\rm NL}$ type (see \S \ref{sec:nongauss}), 
and we will examine how the consistency relations are violated in such a case. 
The ultimate goal would be to check consistency relations in observational data, and put a bound on local $f_{\rm NL}$ for instance. The robustness of the consistency relations means we can freely employ data in the highly nonlinear regime (the high momentum $\vec k$ modes), involving astrophysically realistic fluctuations, e.g. galaxies.

\vspace{0.5em} 

\noindent 7. One might worry that the consistency relation could be
violated by the finite size of the simulation box, especially
for the symmetry transformation that
involves shifting the gravitational potential by a term linear in $\vec x$
(Eq. \eqref{lineargradientSymm}), which seems na\"ively inconsistent
with the periodic boundary condition of the simulations.
However, from the point of view of the particles, all they see is
the gradient of the potential, and the symmetry in question
simply shifts this gradient by a constant, which does respect the
periodic boundary condition.
The fact that, as we will see, the consistency relations hold in the
$N$-body data indeed confirms this expectation.

\vspace{0.5em}

\noindent 8. Lastly, it should be kept in mind that in the presence of features in the power spectrum (e.g. acoustic peaks), the bispectrum could present a $1/q$ behavior for mildly squeezed triangles, albeit recovering the behavior expected from Eq.~\eqref{nopoles} in the strict $q\to0$ limit.\footnote{We are grateful to Marko Simonovi\'c for pointing this out.} When such features are present a simple power series in $q$ will not suffice to describe the squeezed bispectrum, and the complete dependence should be taken into account~\cite{Mirbabayi:2014gda,Baldauf:2015xfa}. For the range of $k$'s we are considering here, this is a negligible effect any way.

\vspace{0.5em}

To summarize, the goal of this paper is twofold. First, we test the
consistency relations \eqref{nopoles}  at equal time 
using the results of $N$-body simulations with gaussian initial
conditions, focusing on the three-point function or bispectrum.
To the best of our knowledge, this is the first time that the
consistency relations have been verified 
for scales that are well within the non-perturbative
regime.\footnote{\label{otherCRs} A different kind of consistency relation has been tested
  in~\cite{Nishimichi:2014jna} using $N$-body simulations as
  well. That interesting relation 
  concerns the higher order coefficients of
  the low $q$ expansion of the bispectrum
  \cite{Kehagias:2013paa,Valageas:2013zda}. 
  More specifically, it
  concerns the $q^0$
  behavior in the context of (\ref{nopoles}), and its derivation
  crucially rely on the hard observables being mass fluctuations as
  opposed to galaxy fluctuations. The
  consistency relations we focus on 
  are instead more robust and valid even for galaxy
  observables.} Secondly, we show that when the initial conditions
for the primordial fields are non-gaussian of the local $f_{\rm NL}$
type, deviations from \eqref{nopoles} are observed, 
as expected from theoretical
arguments~\cite{Peloso:2013zw,Horn:2014rta,Valageas:2016hhr}.

\section{Checking the consistency relations in $N$-body simulations}
\label{NbodyCheck}

We describe in \S \ref{sec:setup} our methodology, 
focusing in particular on how to obtain the
bispectrum in the squeezed limit. This is followed by a discussion 
in \S \ref{sec:fit} of
how we fit the bispectrum with a power series in the squeezed momentum $q$.
The results of the fit are presented in \S \ref{sec:results}, for
$N$-body simulations with gaussian initial conditions. 
We verify that the consistency relations are indeed satisfied, even
though the high momentum modes are in the non-linear regime. We draw
attention to, and comment on, the fact that the linear-gradient consistency relation
(i.e. the lack of $1/q$ pole in \eqref{nopoles}) is satisfied,
despite the periodic boundary
conditions of the simulations---which one might na\"ively expect to invalidate the linear
gradient symmetry of Eq.~\eqref{lineargradientSymm}.
We demonstrate in \S \ref{sec:nongauss} that the consistency relations are violated 
for simulations with non-gaussian initial conditions of the local $f_{\rm NL}$ type.

\subsection{Setup and details of the measurement} \label{sec:setup}

We use a suite of $N$-body simulations consisting of $N_r = 40$
realizations with gaussian initial conditions. The box size is 
$L = 2.4$ Gpc/$h$ comoving, with $1280^3$ particles. The cosmological
parameters are $\Omega_\Lambda=0.75$, $\Omega_m=0.25$ (of which $\Omega_b=0.04$),
$h=0.7$, $n_s=1$ and $\sigma_8=0.8$. We analyze the simulation outputs at redshift $z=0$.
For further details on the simulations, see~\cite{Scoccimarro:2011pz}.

\begin{figure}[t]
\centering
\includegraphics[width=0.30\textwidth]{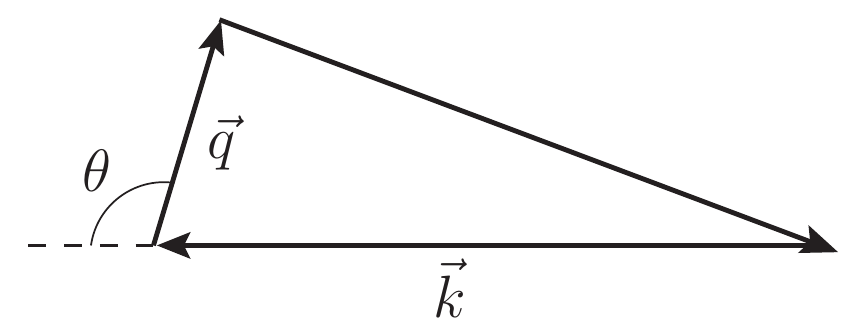}
\caption{The bispectrum is defined for three Fourier modes whose
  momenta sum to zero. The resulting triangle can be uniquely labeled
  by $q \equiv | \vec q \,|$ (the softest momentum), $k \equiv |\vec k|$ (the
  hardest momentum), and the angle between them $\theta$.
By virtue of the fact that $q$ is the softest and $k$ is the hardest, $\theta$
is between $\pi/2$ and $\pi$.}
\label{fig:triangle}
\end{figure}

A prime observable of focus is the bispectrum, in the so-called
squeezed limit, i.e. when one of the legs (in momentum
space) is soft. We are particularly interested in what happens when
that leg, labeled by the momentum $q$, becomes softer and softer 
as other quantities that label the relevant momentum-space triangle are kept
fixed. A convenient parametrization is to take them to be the
highest momentum leg, labeled by $k$, and its angular separation from
the soft leg, labeled by $\theta$. With this choice,
  $\theta$ is between $\pi/2$ and $\pi$ (see Fig.~\ref{fig:triangle}).\footnote{By restricting ourselves to $k$ being the highest momentum
and $\theta$ between $\pi/2$ and $\pi$, we are implicitly assuming
parity invariance: that two triangles related to each other by a
reflection have the same bispectrum.
}
We will have more to say about the choice of parametrization below.

The bispectrum in the squeezed limit can then be expressed as a power
series in the soft mode,
\begin{align} \label{eq:series}
B_\delta(q,k,\theta)=\sum_{n=-2}^\infty a_n(k,\theta)P_\delta(q)\,q^n\,.
\end{align}
We will truncate this power series at some finite $n$, with the
understanding that this is a good approximation for small values of
$q$---the precise $n$ at which we truncate will be determined by the
goodness-of-fit to the data.
The consistency relations~\eqref{nopoles} tell us that, for equal time
correlators, one has $a_{-2}(k,\theta)=a_{-1}(k,\theta)=0$. The goal
of this paper is to check this prediction. We wish to do it in a way
that does not assume any knowledge of the
coefficients $a_n$. They are known robustly only within
perturbation theory, that is, if $k$ is not too large. For large
$k$'s,
non-linearity, or baryonic physics in the case of galaxy
observables (in anticipation of applications to observational data), 
makes it difficult to robustly predict $a_{n}$. Thus we carry
out the analysis without prior assumptions on them.

The key feature we exploit is that Eq.~\eqref{eq:series} takes a 
{\it factorized} form: for each $n$, the dependence on the soft
momentum $q$ is factorized from the dependence on the hard momentum
$k$ (and $\theta$).
The coefficients that contain the $k$ and $\theta$ dependence,
$a_n(k,\theta)$, can be treated as free parameters when fitting the
bispectrum. As a simplifying procedure, 
since we are not ultimately interested in the $k$ and $\theta$
dependence of the bispectrum or $a_n$, we average over all possible values of $k$ and
$\theta$ when we measure the bispectrum for a given $q$.\footnote{We will later check this procedure by varying the range of $\theta$
over which we average.}

At this point, a subtlety occurs because of the discrete
nature of the Fourier modes in a finite volume.
Let us focus on the coefficient $a_n$ for a particular $n$.
Our procedure is effectively to compute some averaged version of $a_n$ by
summing over all possible $k$'s and $\theta$'s at a fixed $q$, i.e.
summing over all triangles which has one momentum
leg of magnitude $q$. The issue is this: within our set of discrete
Fourier or momentum modes, for a given $q$, not all possible $k$'s
and $\theta$'s are actually allowed---in fact, the span of possible
$k$'s and $\theta$'s would depend on the value of $q$ in a subtle way;
this means the averaged $a_n$ would end up inheriting a subtle $q$
dependence. 
{\it This $q$ dependence cannot be predicted without prior
knowledge or assumption of how $a_n$ depends on $k$ and $\theta$.}
It is useful to concretely see how this comes about by dividing
the $k$'s and $\theta$'s into bins, labeled by $i$. For instance,
a bin centered at $(k_i, \theta_i)$ might have contributions from
$N_{i,q}$ triangles. Note how the $q$ dependence is ``sneaked'' in
through the fact that $N_{i,q}$ depends on $q$. In this language,
averaging over all possible triangles for a given $q$ amounts
to computing the following:
\begin{align}
\bar B_\delta(q)&=\frac{\sum_i B_\delta(q,k_i,\theta_i) N_{i,q}}{\sum_i N_{i,q}} \notag \\
&=\sum_{n=-2}^\infty\frac{\sum_i a_n(k_i,\theta_i) N_{i,q}}{\sum_i N_{i,q}} P_\delta(q)q^n \notag \\
&\equiv \sum_{n=-2}^\infty \bar a_n(q)P_\delta(q)q^n\,. \label{eq:Bbad}
\end{align}
We are thus left with an averaged $a_n$, which we call $\bar a_n$,
that has an unwanted $q$ dependence which cannot be predicted without
making assumptions about how $a_n$ behaves for high $k$'s.
Thus, imagine we fit the $N$-body data with $n$ up to, for example, $1$.
Even if one puts aside the possible $q$ dependence of 
$\bar a_{-2}$ and $\bar a_{-1}$ (which for Gaussian initial conditions
are expected to vanish), the unknown $q$ dependence of 
$\bar a_0$ and $\bar a_1$ is problematic.

This way of spelling out the problem also suggests its cure.
The above averaging weighs each $i$-th bin by the number of triangles
in it, $N_{i,q}$. We can instead weigh each bin equally (or for
that matter, use any other weights as long as they do not depend
on $q$):\footnote{In practice, this means when we loop through the
  triangles for a given $q$, we weigh them by $1/N_{i,q}$.}
\begin{align} \label{eq:Bsum}
\bar B_\delta(q)=\frac{\sum_i B_\delta(k_i,\theta_i,q)}{\sum_i}\equiv \sum_{n=-2}^\infty \bar a_nP_\delta(q)q^n\,.
\end{align}
The coefficients $\bar a_n$ are now given by
\begin{align}
\label{anbarcoeff}
\bar a_n\equiv\frac{\sum_i a_n(k_i,\theta_i)}{\sum_i}\,,
\end{align}
and are truly independent of the soft momentum. They
are treated as free parameters in our fit of the data.

To simplify the analysis, we also bin in $q$.
In particular, if $\{ \bar q\}$ is a bin with average soft momentum $\bar q$, the binned version of the bispectrum~\eqref{eq:Bsum} is
\begin{align} \label{eq:fittingmodel}
\bar B_\delta (\bar q)=\sum_{q\in\{\bar q\}} \bar B_\delta (q)=\sum_{n=-2}^{n_\text{max}} \bar a_n M_n(\bar q)\,,
\end{align}
where $M_n(\bar q)=\sum_{q\in\{\bar q\}}P_\delta(q)q^n$ is also
measured from the data to avoid any theoretical bias. 
The value of $n_\text{max}$ is something we have to experiment with:
qualitatively, the more squeezed our triangles are (smaller $q$'s), the
lower is the $n_\text{max}$ we need.
In the next section we will rely on data to find how many $\bar a_n$'s
we need to account for higher order corrections to the small $q$ expansion.

Lastly, let us comment on our bispectrum triangle parametrization,
described in Fig. \ref{fig:triangle}. 
One alternative 
\cite{Lewis:2011au}
is to parametrize the triangle in terms of 
$\vec q$, $\vec k+\vec q/2$ (where $\vec k$ is one of the two high
momentum legs), and the angle between them, say $\beta$.
Assuming invariance under parity (a reflection of
the triangle), the bispectrum should be unchanged under
${\,\rm cos\,}\beta \rightarrow {\,\rm cos\,} (\pi - \beta)$. 
Thus, if $q$ enters into the bispectrum only through ${\,\rm
  cos\,}\beta$, the squeezed bispectrum should contain only even
powers of $q$, as suggested by \cite{Lewis:2011au}.
This appears to be true in some cases but not in others---for
instance, it can be checked in perturbation theory that if the
initial conditions were of the local $f_{\rm NL}$ type, the
squeezed mass bispectrum depends on the transfer function
at the soft-momentum $q$ (which equals $1$ when $q = 0$, but
has corrections with both even and odd powers of $q$)~\cite{Peloso:2013zw}.\footnote{Even in those cases where the squeezed bispectrum 
parametrized according to \cite{Lewis:2011au}
appears as an even-power series in $q$, 
the information does not allow us to for instance
infer $\bar a_{-1}$ from $\bar a_{-2}$ (these are parameters
we are ultimately interested in, using our parametrization).
In those cases, $\bar a_{-1}$ is not directly related to $\bar a_{-2}$ 
but is instead related to the
average over $k$ and $\theta$ of the derivatives of $a_{-2}$ with
respect to $k$ and $\theta$. We thank Antony Lewis for discussions
on different parametrizations of the bispectrum triangle.
}
We thus do not find a significant advantage for using
the alternative parametrization, although the bispectrum can be
analyzed that way if one wishes.

\subsection{Details of the fit} \label{sec:fit}

\begin{figure}[]
\centering
\includegraphics[width=0.48\textwidth]{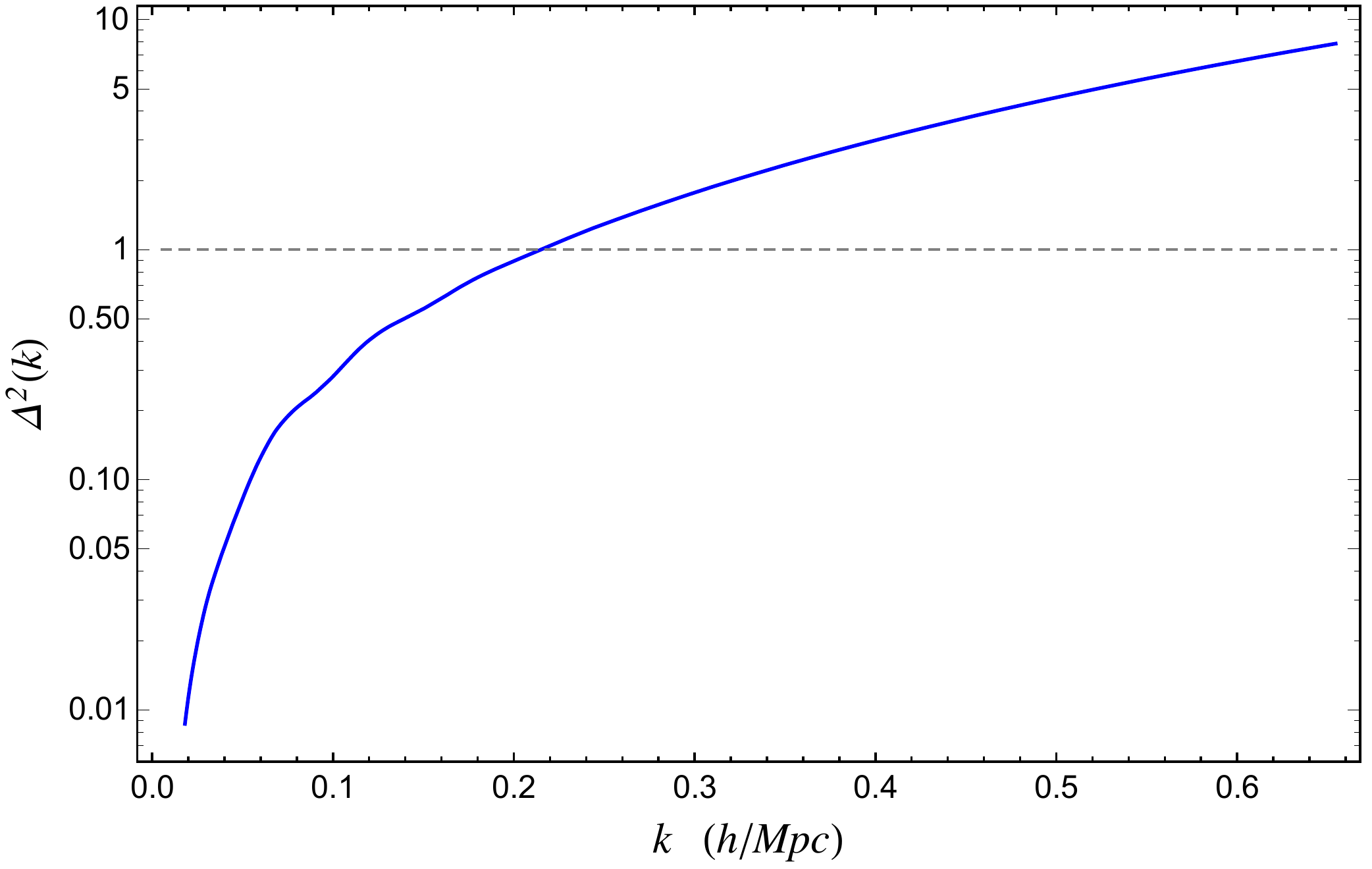}
\caption{Measured value of $\Delta^2(k)=4\pi k^3P_\delta(k)$. For the hard modes under consideration, $k\in[0.52,0.66]$ $h$/Mpc, one notices that $\Delta^2(k)\gtrsim5$, i.e. we are far away from the linear regime.} \label{fig:delta2}
\end{figure}

Our analysis is performed averaging the bispectrum as in
Eq.~\eqref{eq:Bsum} over hard modes ranging from $k=0.52$ $h$/Mpc
to $k=0.65$ $h$/Mpc 
(corresponding to $k=199 k_f$ to
$k=251 k_f$, with $k_f=2\pi/L\simeq2.6\times 10^{-3}$~$h/$Mpc being the
fundamental mode), and over all the relative angles,
$\theta\in[\pi/2,\pi]$\footnote{As a simple check, we also repeated
  our analysis averaging over $\theta\in[\pi/2,3\pi/4]$. 
The conclusions are largely unchanged, suggesting that 
adjusting the angular weighting does not have a significant impact on
the outcome, and that the vanishing of $\bar a_{-2}$ and $\bar a_{-1}$ for
gaussian initial conditions is not the result of accidental
cancellation when averaging over angles.}. As one can see from
Fig.~\ref{fig:delta2} the hard modes we are considering are well within the
non-linear regime. 

We also measure the bispectrum for soft momenta
ranging from $q=3k_f$ to $q=19k_f$, with a bin size $\Delta k=2k_f$. 
The choice of this window for $q$'s requires some explanations.
The high end $q=19 k_f$ is chosen to include as many modes as 
possible (thus minimizing error bars on $\bar a_{-2}$ and $\bar
a_{-1}$), while still staying within the squeezed limit such that the
expansion in Eq.~\eqref{eq:Bsum}, truncated at $n$ of a
few, is a good approximation.

The low end $q = 3 k_f$ is chosen because the procedure of 
eliminating the unwanted $q$ dependence in $\bar a_n$, described in
Sec. \ref{sec:setup}, is actually not perfect. Recall that we form
bins, labeled as $(k_i , \theta_i)$, and compute $\bar B_\delta$ and
$\bar a_n$ (Eqs. \eqref{eq:Bsum} and \eqref{anbarcoeff}) by giving these
bins equal weights. In doing so, it is important that each bin is actually
not empty, that there are triangles that fall into them.
Thus, the bins have to be sufficiently wide. But there is some tension
between using wide bins and using the bin-averaged $a_n$ as a fair
representation of how $a_n$ varies with $k$ and $\theta$.\footnote{In other words, within a wide bin, the precise set of $k$'s and
  $\theta$'s that fall into that bin would depend on $q$, and thus we
  are not achieving the goal of eliminating the unwanted $q$
  dependence.}
In Appendix ~\ref{app:firstbin}, we show a test of our procedure for a
particular model of $a_n (k,\theta)$ (one motivated by perturbation
theory), and check to what extent our procedure yields $\bar a_n$ that
is truly $q$ independent. We find that this works well as long
as $q\geq 3k_f$. Hence we restrict our analysis only
to soft modes such that $q\geq3k_f$.

To determine the best-fit values for the parameters $\bar a_n$ we maximize the following likelihood for each realization, $r=1,\dots,N_r$:
\begin{align} \label{eq:L}
\mathcal{L}_{(r)}\propto\frac{1}{\sqrt{\det \bm{C} }}\exp\left[-\frac{1}{2}\bm{\Delta}^{(r)}\cdot {\bm{C}}^{-1}\cdot\bm{\Delta}^{(r)}\right]\,,
\end{align}
We define the vector $\bm{\Delta}^{(r)}=\bar{\bm{B}}_\delta^{(r)}-\sum_{n}\bar a^{(r)}_n \bar{\bm M}^{(r)}_{n}$,
and the covariance matrix $C_{ij} =\langle \Delta_i\,\Delta_j\rangle$.
All vectors run over the soft momenta, $f_i\equiv f(\bar q_i)$, and
the angular brackets stand for an average over the available
realizations, i.e. $\langle f_i \rangle = \frac{1}{N_r-1}\sum_r
f_i^{(r)}$ (for instance, the covariance matrix is obtained by
averaging over realizations).

Note that the vector $\bm{\Delta}$ depends on the fit
parameters $\bar a_n$, and so the covariance matrix itself depends on the
parameters. We use an iterative procedure (akin to
the Newton-Raphson algorithm) 
to determine the optimal $\bar a_n$'s that maximize the likelihood.
First, we determine $C_{ij}$ with the $\bar a_n$'s set to zero. The
maximization of the likelihood can thus be done analytically, because
the remaining dependence on $\bar a_n$ shows up only in the exponent 
of the likelihood in the standard $\chi^2$
fashion (essentially equivalent to fitting the slope of a straight line).
The resulting best-fit $\bar a_n$'s are plugged back into 
the definition of $C_{ij}$, and the whole procedure is repeated again
to obtain a new set of best-fit $\bar a_n$'s. So on and so forth
until convergence is achieved.

Once this is done, the final value of the ML estimators and their uncertainties is computed from the average and variance over realizations, i.e.
\begin{subequations}
\begin{align}
\bar a_n&=\frac{1}{N_r}\sum_{r=1}^{N_r} \bar a_n^{(r)}\,, \\ 
\sigma_{\bar a_n}^2&=\frac{1}{N_r(N_r-1)}\sum_{r=1}^{N_r}(\bar a_n^{(r)}-\bar a_n)^2\,.
\end{align}
\end{subequations}
Note that the likelihood analysis itself, applied to each realization,
does yield an error estimate, but we deem $\sigma_{\bar a_n}^2$
estimated from the spread between {\it independent} realizations as more
reliable. For one thing, the likelihood analysis treats
  the data vector as Gaussian distributed, which is an approximation.
The desire to have an accurate error estimate is why we analyze 
the realizations one at a time, as opposed to using all of them in one
go.\footnote{Analyzing the realizations all at once 
would give us essentially the same final best-fit $\bar a_n$,
but would not let us reliably estimate the associated errorbar.}

To determine the goodness of the fit, we rely on the Bayesian information criterion (BIC)~\cite{schwarz1978estimating,Liddle:2004nh}:
\begin{align}
\text{BIC}=-2\log\mathcal{L}_\text{max}+N_\text{par}\log N_q\,,
\end{align}
which has been shown to be dimensionally consistent, i.e. not to favor
overfitted models~\cite{Liddle:2004nh}. Here $\mathcal{L}_\text{max}$
is the maximum likelihood combining all realizations, $N_\text{par}$
the number of parameters of the model and $N_q$ the number of data
points used. A model with the lowest BIC represents the best
compromise between maximizing likelihood and minimizing the number of
parameters.

\subsection{Results} \label{sec:results}

\begin{figure*}[t] 
\centering
\includegraphics[width=0.481\textwidth]{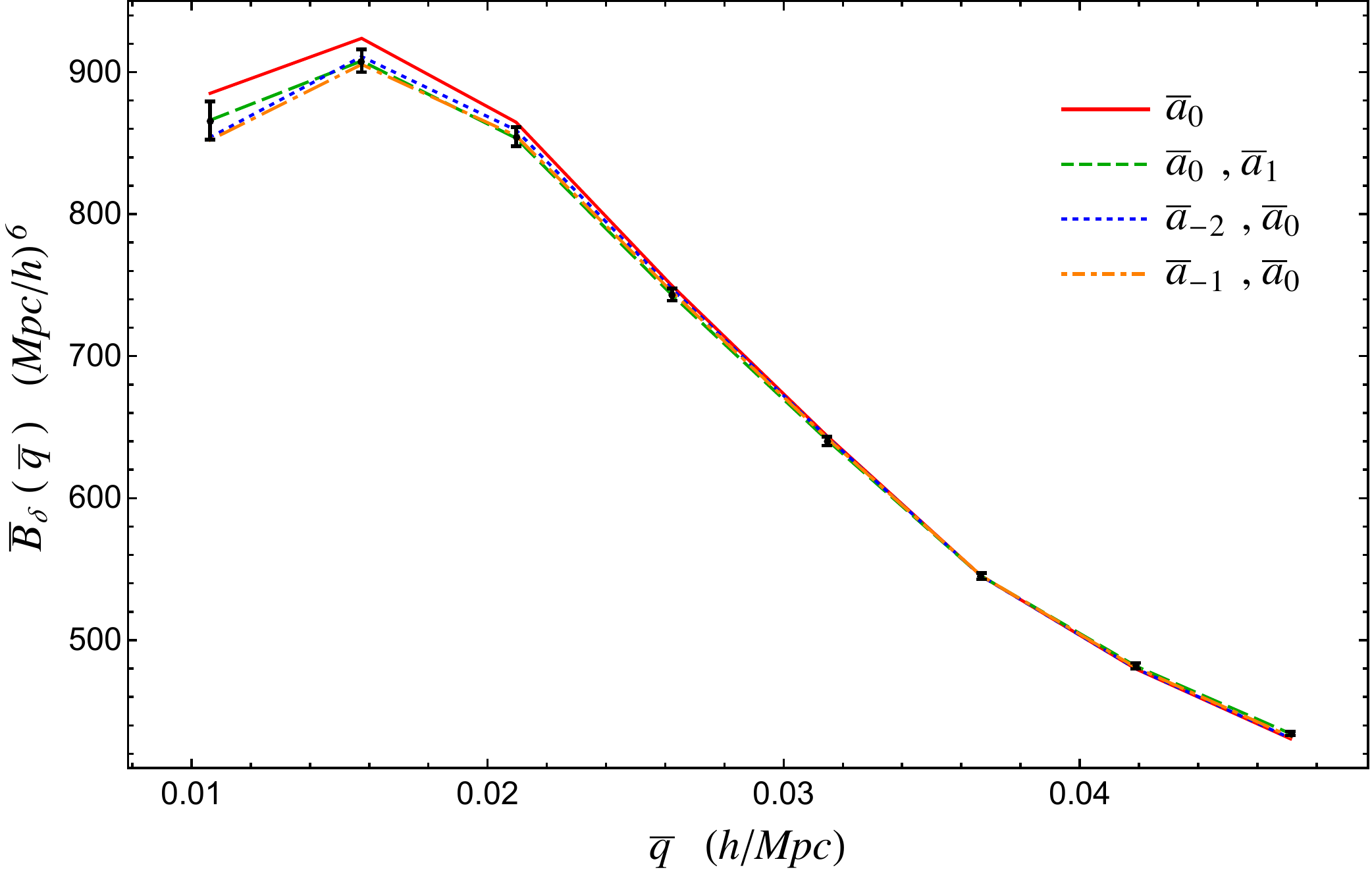} \hspace{0.2em}
\includegraphics[width=0.48\textwidth]{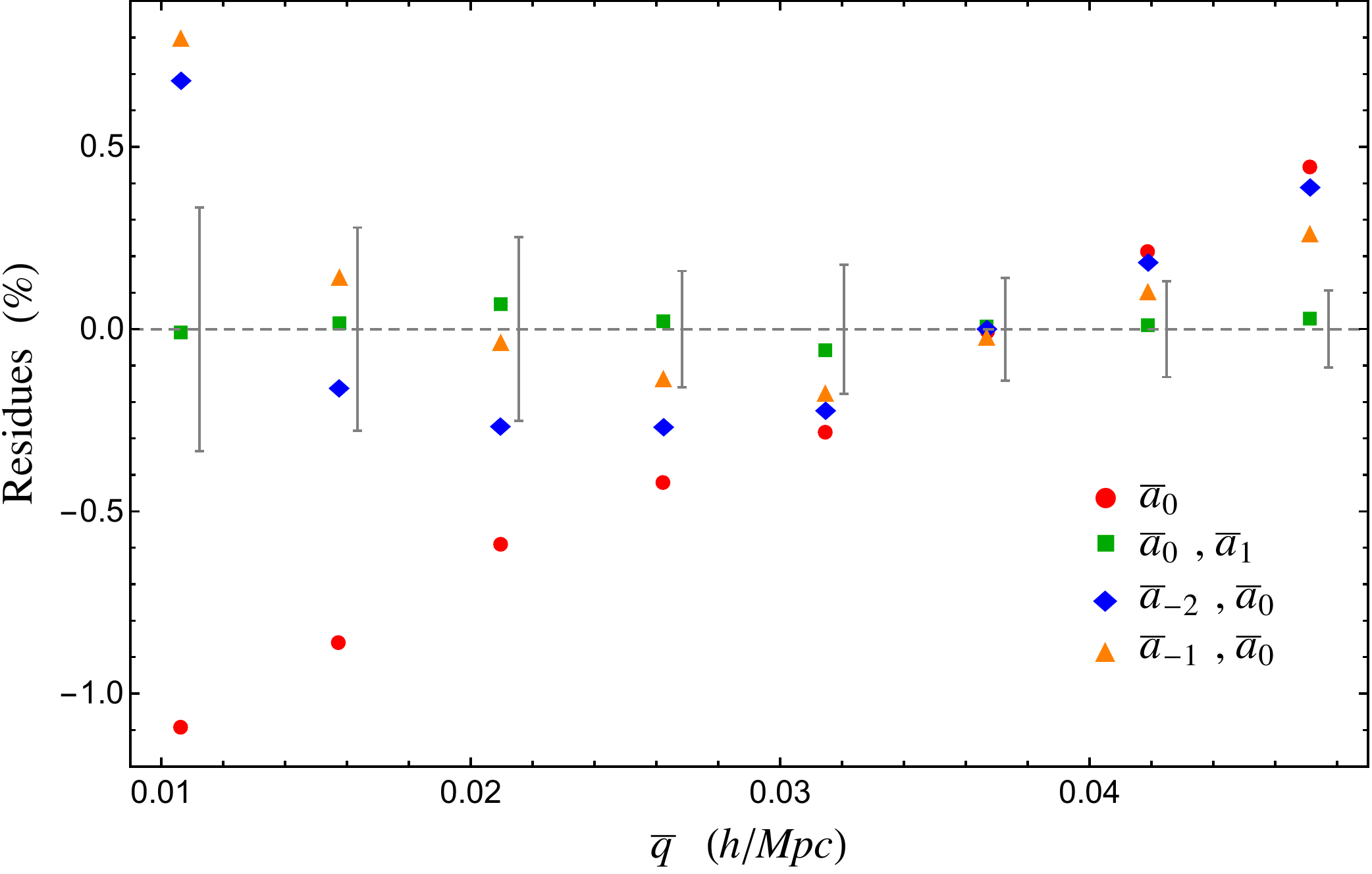}
\caption{Left panel: comparison between $N$-body data (points with errorbars)
  and some of the models (lines with different styles/colors),
  including different sets of parameters. The $y$-axis is 
  the bispectrum $\bar B_\delta$ 
  while the $x$-axis is soft momentum $\bar q$.
Note that $\bar B_\delta$ is a function only of $\bar q$, because we
  have already summed over $k$'s and $\theta$'s (see Eqs.~\eqref{eq:Bsum}
  and \eqref{eq:fittingmodel}).  Here, the errorbars reflect the
  statistical spread in $\bar B_\delta$. 
Right panel: percentage difference
  between data and model for the same set of models as the left panel,
  i.e. the $y$-axis is $[\bar B_\delta (\bar q) - \sum_n \bar a_n
  M_n(\bar q)]/[\bar B_\delta (\bar q) + \sum_n \bar a_n
  M_n(\bar q)]]$, expressed in percentage. The errorbars shown represent
the statistical spread in this residue. The errorbars are model dependent---the
ones shown correspond to that of the $(\bar a_0 , \bar a_1)$ model.
Note that the errorbars are correlated across different $\bar q$'s,
which partly explains why the $(\bar a_0, \bar a_1)$ model appears
well within the errorbars at all momenta.
} \label{fig:fit}
\end{figure*}

Let us now present the results of our analysis. In Fig. ~\ref{fig:fit} we report some of the fits including different sets of parameters as well as the corresponding residues. In Table~\ref{tab:fit} we compare all the models we have tested.
\bgroup
\def\arraystretch{1.5}
\begin{table}[t]
\centering
\begin{tabular}{c | c | c | c}
$\bar a_n$ included & $\bar a_{-2}\, (10^{-6}\text{Mpc}/h)$ & $\bar a_{-1}\, (10^{-2}\text{Mpc}/h)^2$ & BIC \\
\hline\hline
$\bar a_0$ & $-$ & $-$ & 99.63 \\
\hline
$\bar a_0,\bar a_1$ & $-$ & $-$ & 17.77 \\
\hline
$\bar a_0,\bar a_1,\bar a_2$ & $-$ & $-$ & 19.82 \\
\hline
$\bar a_{-2},\bar a_0$ & $-30.4\pm5.1$ & $-$ & 65.50 \\
\hline
$\bar a_{-2},\bar a_0,\bar a_{1}$ & $0.2\pm6.7$ & $-$ & 19.84 \\
\hline 
$\bar a_{-1}, \bar a_{0}$ & $-$ & $-42.3\pm5.4$ & 39.57 \\
\hline
$\bar a_{-1},\bar a_{0},\bar a_{1}$ & $-$ & $0.6\pm10.3$ & 19.84 \\
\hline
$\bar a_{-2},\bar a_{-1}, \bar a_{0}$ & $69\pm16$ & $111\pm17$ & 22.35 \\
\hline
$\bar a_{-2},\bar a_{-1},\bar a_{0},\bar a_{1}$ & $16\pm55$ & $26\pm87$ & 21.83 \\
\hline\hline 
\end{tabular}
\caption{Detailed results of the likelihood fits with different sets of
  parameters ($\bar a_n$'s). The Bayesian information criterion
  clearly selects the $(\bar a_0, \bar a_1)$ model.} \label{tab:fit}
\end{table}
\egroup

Focus first on the first three models in the table, which do not
involve $\bar a_{-2}$ nor $\bar a_{-1}$. 
We see that the first model, involving $\bar a_0$ alone, is not
a good fit to the $N$-body data (from both the BIC value in the table
and from Fig.~\ref{fig:fit}). Adding $\bar a_1$ greatly improves
the fit, while further adding $\bar a_2$ does not lower the BIC score.
Recall that in our power-series fit of the squeezed bispectrum as a function
of a range of soft momenta (Eq.~\eqref{eq:fittingmodel}), we do not
know a priori how many higher order terms we need. This exercise tells us
it is sufficient to stop at $\bar a_1$ (but necessary to include it), 
with the kind of precision and the range of soft momenta we have.

The rest of the models in the table involve $\bar a_{-1}$ and/or $\bar
a_{-2}$. In all cases, the BIC score worsens. The inferred values
for $\bar a_{-1}$ and $\bar a_{-2}$ are consistent with zero, except
for the $(\bar a_{-2}, \bar a_{-1}, \bar a_0)$ model. For this model,
the fit prefers non-zero values for $\bar a_{-2}$ and $\bar a_{-1}$ 
to compensate for the lack of a $\bar a_1$ term. Note however this
model has a worse BIC score compared to the $(\bar a_0, \bar
a_1)$ model. It is also reassuring that the $(\bar a_0, \bar a_1)$
fits the data well, with residues that have no clear trend with
momenta (see the right panel of Fig. \ref{fig:fit}). 

We conclude from this exercise that the $N$-body data, with gaussian
initial conditions, are consistent with a vanishing value for $\bar a_{-2}$ and for $\bar a_{-1}$, confirming
expectations from the consistency relations.

\subsection{Violation of the consistency relations from non-gaussian
  initial conditions}
\label{sec:nongauss}

\begin{figure*}[t]
\centering
\includegraphics[width=0.475\textwidth]{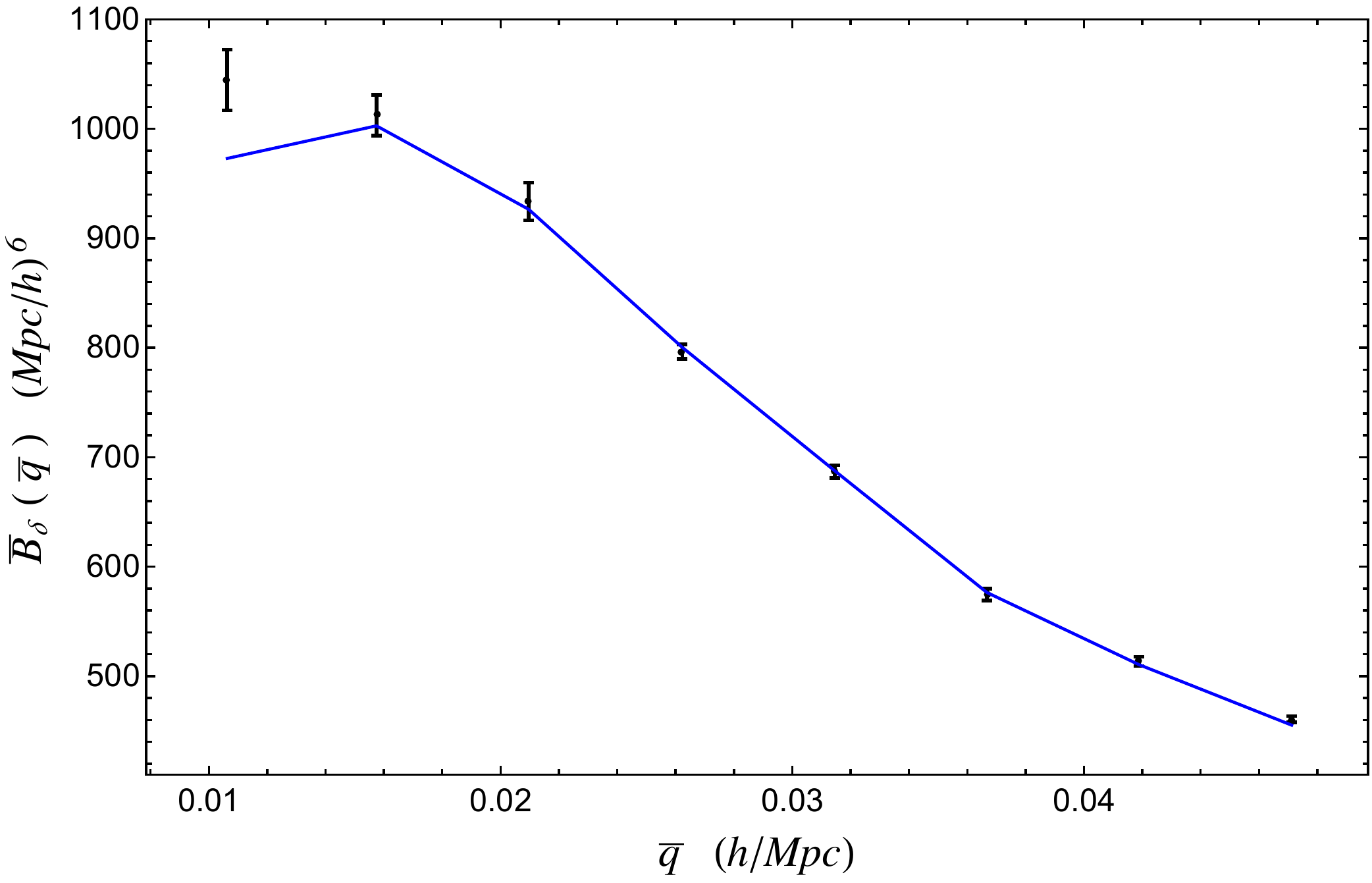} 
\includegraphics[width=0.472\textwidth]{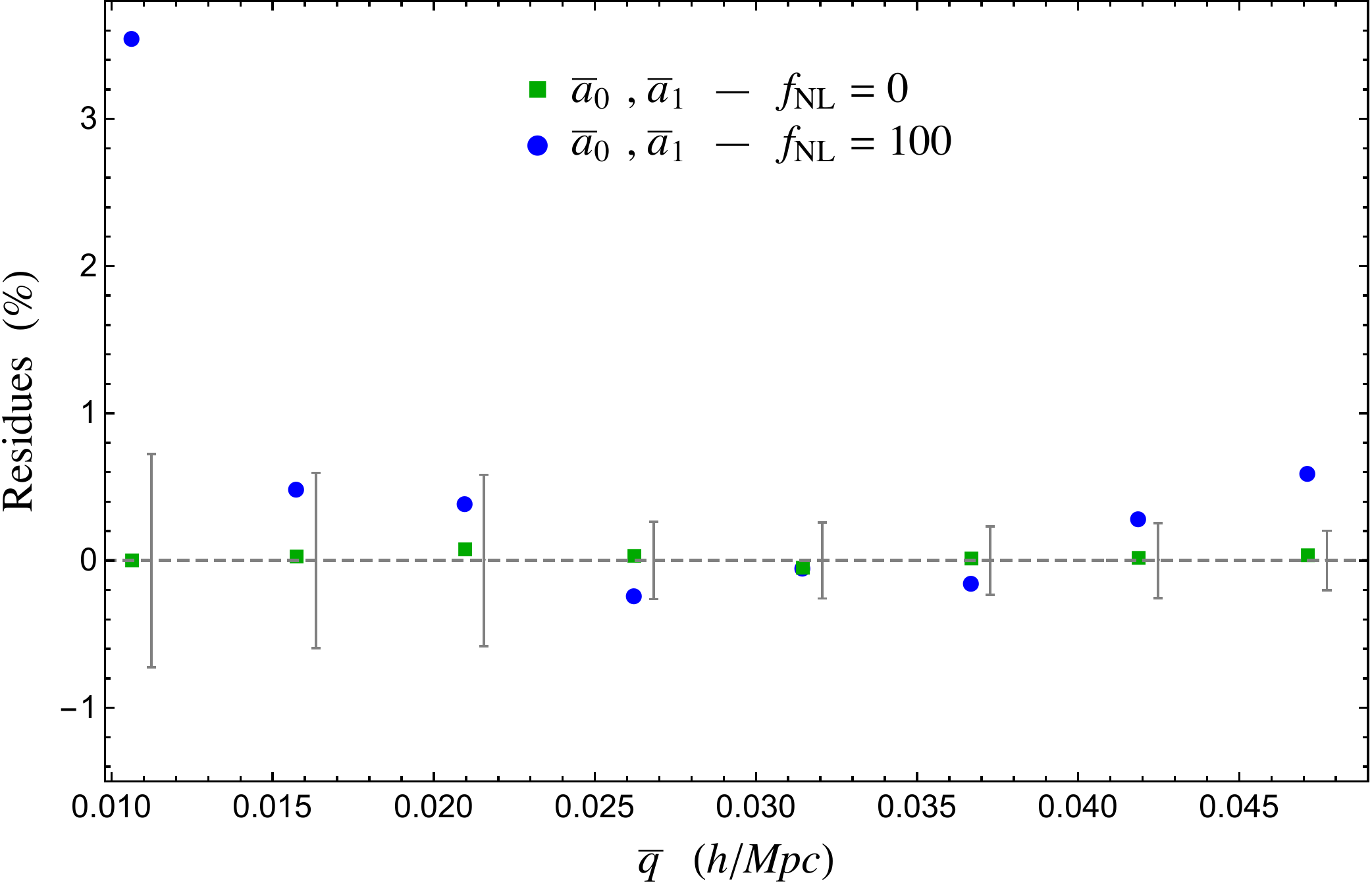}
\caption{Left panel: comparison between $N$-body data and a $(\bar a_0,
  \bar a_1)$ model fit for a case where the initial conditions are
  non-gaussian (of the local $f_{\rm NL}$
type, with $f_{\rm NL} = 100$). Right panel: the blue circles represent the residues
for the ($\bar a_0$, $\bar a_1$) model i.e. $[\bar B_\delta (\bar q) - \sum_n \bar a_n
  M_n(\bar q)]/[\bar B_\delta (\bar q) + \sum_n \bar a_n
  M_n(\bar q)]$ expressed in percentage, where
  $\bar B_\delta$ is the $N$-body bispectrum with initial conditions of
  local $f_{\rm NL} = 100$. The errorbars shown represent the
  statistical spread in this residue. Also shown as green squares are the same
  residues as the green squares in Fig. \ref{fig:fit}, i.e. residues
  for the gaussian case ($f_{\rm NL} = 0$).
  The larger errorbars compared to those in Fig. \ref{fig:fit}
  reflect the fact that fewer realizations are used in this analysis.
} \label{fig:fnl100}
\end{figure*}

In this section we show that, when the initial conditions for the
cosmological fields are non-gaussian (of the local $f_{\rm NL}$ type), 
statistically significant 
deviations from the consistency relations in Eq.~\eqref{nopoles} 
are observed.\footnote{Specifically, the local $f_{\rm NL}$ model is this:
the primordial Bardeen potential $\Phi_p (\vec x) = \phi (\vec x) +
f_{\rm NL} (\phi (\vec x)^2 -
\langle \phi^2 \rangle)$, where $\phi$ is a gaussian random field.
The Bardeen potential (after multiplication
by the transfer function) gives the gravitational potential
used in initializing $N$-body simulations
(see \cite{Scoccimarro:2011pz} for details).
A primordial non-gaussianity of this type is motivated by the curvaton
and modulated reheating models
\cite{Lyth:2006gd,Dvali:2003em,Kofman:2003nx}.
}

We employ a smaller set of $N_r=12$ realizations with the same
cosmological parameters as before, but with an initial matter
distribution characterized by a local non-gaussian parameter
$f_\text{NL} = 100$. The details of the measurement and the analysis
are the same as in Sections~\ref{sec:setup} and \ref{sec:fit}. For the
sake of checking whether or not deviations from consistency relations
occur we limit our analysis to models that only include $\bar a_0$ and
$\bar a_1$ (in addition to possibly $\bar a_{-2}$ and $\bar a_{-1}$). 
We leave a more detailed study of the realizations with non-gaussian
initial condition for future work~\cite{toappear}.

The result of our analysis is unambiguous. The model with just $\bar
a_0$ and $\bar a_1$ (i.e. no poles in the soft limit), which fits very
well the bispectrum in the case of gaussian initial conditions, is not
a good description of the data obtained from $f_\text{NL}=100$. From
the fit we obtain $\text{BIC}=74.12$, much larger than the one
reported in Table~\ref{tab:fit} for the gaussian case. 
Moreover, Fig.~\ref{fig:fnl100} shows that the likelihood fit for this model is not a good description of the data, which is also confirmed by the fact that the residues exhibit a parabolic pattern around zero.
Indeed, introducing either $\bar a_{-2}$ or $\bar a_{-1}$ to the fit
one obtains values that are statistically different from zero: $\bar
a_{-2} = (11.3\pm1.8) \times 10^{-5}$~Mpc/$h$ with BIC$=30.39$, or
$\bar a_{-1}=(16.5\pm2.8)\times 10^{-3}$~(Mpc/$h$)$^2$ with
BIC$=36.79$.\footnote{
For completeness, let us mention several additional models we
investigated: 
the $(\bar a_0, \bar a_1, \bar a_2)$ model has
a BIC of $56.5$, and 
the $(\bar a_{-2}, \bar a_0, \bar a_1, \bar a_2)$/
$(\bar a_{-1}, \bar a_0, \bar a_1, \bar a_2)$/
$(\bar a_{-2}, \bar a_{-1}, \bar a_0, \bar a_1, \bar a_2)$/
$(\bar a_{-2}, \bar a_{-1}, \bar a_0,
\bar a_1)$ models have respectively a BIC score of
$31.8, 34.3, 32.2, 31.2$. 
}
The inferred values for $\bar a_{-2}$ and $\bar a_{-1}$ can in
principle be turned into an estimate of $f_{\rm NL}$, which we leave for
future work.

This shows that, in presence of a non-gaussian distribution 
(of the local $f_{\rm NL}$ type) for the initial cosmological fields, the consistency relations in Eq.~\eqref{nopoles} are violated as expected~\cite{Peloso:2013zw,Horn:2014rta,Valageas:2016hhr}.

\section{Discussion}

The search for primordial non-gaussianities has been so far a 
challenging task.
This is partly due to our lack of theoretical control over observables that are outside the linear regime. Consistency relations are non-perturbative statements that follow solely from symmetry arguments and, as such, might provide a key tool to overcome these difficulties.

In this paper we successfully test them, for the first time, in a
regime well outside the domain of perturbation theory. In doing so, we
highlight and solve a number of technical and conceptual subtleties
associated with the analysis of the bispectrum in the squeezed regime from $N$-body simulations,
whose systematic study has been lacking from the literature
(see \cite{Nishimichi:2014jna} for an exception, though see
footnote \ref{otherCRs}).

Moreover, we show how in presence of non-gaussian initial conditions
of the local $f_{\rm NL}$ type, significant deviations from the
standard consistency relations are observed. 
This is the first step towards extracting constraints on $f_{\rm NL}$
from observational data, using the consistency relations (or
violations thereof). The appeal is that with this method, (non-linear) modes that
are normally discarded can now be used.
Several issues need to be investigated before this goal
can be realized. They include: checking the consistency relations
(1) for biased observables such as halos in $N$-body simulations
or galaxies in hydrodynamic simulations, and (2) including redshift
space distortions.
As explained in Sec. \ref{intro}, the consistency relations are
expected to be robust against these complications, but it would be
useful to test the expectations against simulations---our simple
exercise presented in this paper suggests there could well be
subtleties that need to be understood and addressed.

\begin{acknowledgments}
The authors are grateful to A.~Joyce, A.~Lewis, A.~Nicolis, R.~Penco,
M.~Pietroni, M.~Simonovi\'c and
S.~Wong for interesting and useful discussions, and to M.~Abitbol for
illuminating insights on the statistical analysis. The work done by
A.E. is supported by the Swiss National Science Foundation under
contract 200020-169696 and through the National Center of Competence
in Research SwissMAP. The work done by L.H. is supported in part by
the NASA grant NXX16AB27G and the DOE grant DE-SC011941.
\end{acknowledgments}

\appendix

\section{Checking soft momentum factorization} \label{app:firstbin}

\begin{figure*}[tb]
\centering
\includegraphics[width=0.47\textwidth]{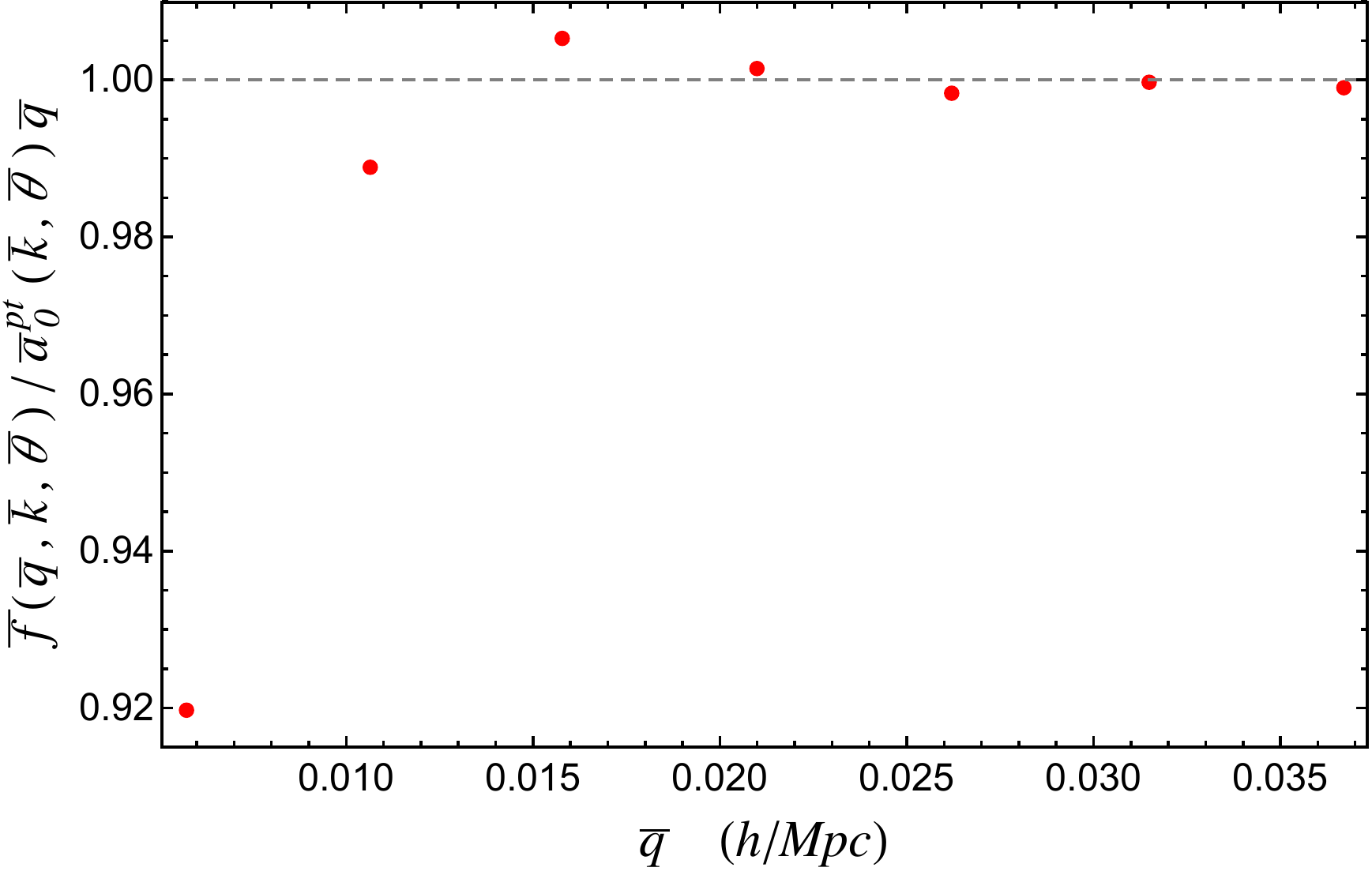}
\caption{Comparison between $\bar f(\bar q)$ and $\bar
  a_0^\text{pt}\,\bar q$. The procedure outlined around Eqs.~\eqref{eq:Bsum} and \eqref{anbarcoeff}, if it works, should make the
two very close to each other. (Here, $\bar f$ is the analog of $\bar
B_\delta$ over there.)
One can notice that, for very small soft momenta, the
procedure does not work so well.} \label{fig:factorization}
\end{figure*}

In this Appendix we show that the procedure outlined around
Eqs.~\eqref{eq:Bsum} and \eqref{anbarcoeff} might not eliminate
the unwanted $q$ dependence if $q$ is extremely small.

As an explicit check let us consider the result obtained in perturbation theory. When the hard mode $k$ is within the linear regime one can easily show that the squeezed limit of the bispectrum (see e.g.~\cite{Bernardeau:2001qr,Horn:2014rta}) gives
\begin{align}
\begin{split}
a_0^\text{pt}(k,\theta)&=\left( \frac{13}{7}+\frac{8}{7}\cos^2\theta \right)P_\delta(k)-\cos^2\theta \, k \,P_\delta^\prime(k)\,.\notag
\end{split}
\end{align}
Let us then consider the toy model
$f(q,k,\theta)=a_0^\text{pt}(k,\theta)q$
(i.e. $f$ is our toy bispectrum for which $a_0^\text{pt}$ as a
function of $k$ and $\theta$ is exactly known), and bin it as in
Eqs.~\eqref{eq:Bsum} and \eqref{eq:fittingmodel} over $k\in[79,91]k_f$
and all relative angles. Let us call the result $\bar f$. Recall that the
worry was that, with a discrete set of triangles, the average over $k$
and $\theta$ of $a_0^\text{pt}$ (we call this $\bar a_0^\text{pt}$) would secretly depend on $q$. 
In our toy example, since the $k$ and $\theta$ dependence of
$a_0^\text{pt}$ are precisely known, we can compute this average
exactly without reference to the particular triangles we happen to
have in our discrete grid. If our procedure to remove the unwanted $q$ dependence
works, then it should be $\bar f(\bar q)=\bar a_0^\text{pt} \bar q$.

In Figture~\ref{fig:factorization} we report the result of our
measurement. As one can see, the procedure works reasonably well only
if the soft momentum is $q\gtrsim3 k_f$. 
In the analysis reported in Section~\ref{sec:results} we therefore
exclude the lowest momentum bin. See Sec. \ref{sec:fit} for a
discussion of why our procedure does not perfectly remove the unwanted
$q$ dependence.

Finally, it can be checked that, in presence of local $f_\text{NL}$ non-gaussianities, the factorization in Eq.~\eqref{eq:fittingmodel} holds better, even for low momenta. Indeed, the dominant $a_{-2}(k)$ term is expected to have no $\theta$-dependence and only a mild $k$-dependence~\cite{Peloso:2013zw}.

\bibliographystyle{apsrev4-1}
\bibliography{biblio}

\end{document}